\documentstyle[editedvolume]{crckapb} 

\def\plotfiddle#1#2#3#4#5#6#7{\vbox to #2{
\includegraphics{#1}}}

\begin{opening}
\title{High Resolution WFPC2 Imaging of IRAS 09104$+$4109}

\author{L. Armus}
\institute{SIRTF Science Center, California Institute of Technology, MS 310-6, 
Pasadena, CA 91125}

\author{B. T. Soifer}
\author{G. Neugebauer}
\institute{California Institute of Technology, MS 320-47, Pasadena, CA 91125}

\end{opening}

\runningtitle{WFPC2 Imaging of IRAS 09104$+$4109}

\begin{document}


\section{Introduction}

With an infrared luminosity of L$_{IR}=6\times10^{12}$L$_{\odot}$, IRAS
09104$+$4109 is one of the most luminous objects detected in the IRAS
survey.  First identified with a faint, R$\sim$18 mag, galaxy at $z=0.442$
by Kleinmann {\it et al.}(1988), IRAS 09104$+$4109 emits more than $99\%$
of its bolometric luminosity longward of $\lambda \sim1\mu$m.  Like most
Ultraluminous Infrared Galaxies (ULIRGs), IRAS 09104$+$4109 has very
strong emission lines.  The nucleus exhibits a Seyfert type 2 spectrum,
with strong, narrow lines over a wide range of ionization (Kleinmann {\it
et al.},1988; Soifer {\it et al.},1996).  Two--dimensional, ground--based
spectrophotometry shows a ``plume" of [OIII] emission--line gas extending
almost $5''$ from the nucleus, along with a strong, blueshifted
($\Delta$V$\sim1200$ km s$^{-1}$) nuclear component \cite{cv96}.  Although
the dominant energy source in most ULIRGs is still a matter of some
debate, the highly polarized spectrum ($\sim18\%$), broad ($10^{4}$ km
s$^{-1}$), (presumably reflected) MgII emission line, and double--lobed,
core--jet radio structure, indicate a quasar at the heart of IRAS
09104$+$4109 \cite{hw93}.  Recent calculations by Hines {\it et al.}
(1999) suggest that the buried quasar in IRAS 09104$+$4109 can account for
all of the bolometric luminosity of the system.

What makes IRAS 09104$+$4109 unique among ULIRGs, is its environment.
While nearly all known ULIRGs are either pairs of interacting, spiral
galaxies or post-merger systems, (Armus {\it et al.}, 1987; Sanders {\it
et al.}, 1988; Lawrence {\it et al.}, 1989; Murphy {\it et al.}, 1996),
IRAS 09104$+$4109 is a cD galaxy at the center of a rich cluster
(Kleinmann {\it et al.}, 1988; Hutchings \& Neff 1988).  The apparent
shape of this cluster is relatively flat (a/b$\sim0.5 - 0.7$), suggesting
a dynamically unrelaxed system.  The cD halo of IRAS 09104$+$4109 itself
extends at least $25-30$ kpc in radius, with a number of ``knots" seen in
projection, having $21<$R$<24$ mag and $3<$R$-$K$<5$ mag \cite{s96}.  The
IRAS 09104$+$4109 cluster also contains a large reservoir of hot,
inter--galactic gas.  X-ray observations with the ROSAT and ASCA
satellites (Fabian \& Crawford 1995) show a resolved source extending over
$30''-40''$ around IRAS 09104$+$4109, but having a central ``hole" in the
emission.  The X-ray morphology, and particularly the X-ray spectra,
suggest that IRAS 09104$+$4109 sits at the center of a cooling flow with a
mass deposition rate of as much as $500-1000$ M$_{\odot}$ yr$^{-1}$.

There are a number of fundamental, un--answered questions concerning IRAS
09104$+$4109 which we might hope to address with high--resolution,
multi--color imaging.  First, what is the source of the gas and dust ?  Is
it more likely to be the cluster (i.e. the cooling flow), or cannibalized
cluster members ?  Second, what is the structure of the ionized gas and
the stellar continuum on kpc scales ?  Can the morphology provide clues as
to the excitation and possibly the kinematics in the circumnuclear
environment (e.g. sub--structures such as radiating filaments or blue
clusters that suggest reflected nuclear light or young stars) ?  Finally,
how do the small (sub--kpc) and large--scale features relate to the radio
jets and the cluster as a whole ?  In order to address these questions, we
have obtained high--resolution, WFPC2 images of IRAS 09104$+$4109 with the
Hubble Space Telescope.  The results of these imaging observations are
described here.  Note, that at the redshift of IRAS 09104$+$4109, $1.0''$
corresponds to a projected linear dimension of 5.07 kpc, for an H$_{0}=75$
km s$^{-1}$ Mpc$^{-1}$ and q$_{0}=0$.

\section{Observations and Data Reductions}

IRAS 09104$+$4109 was observed with the WFPC2 in the F622W, F814W, and
FR680N filters.  The F622W and F814W filters are broad--band filters with
effective wavelengths of $\sim6200$\AA~ and  $\sim8000$\AA~,
respectively.  The FR680N filter is a $\sim1.3\%$ bandpass linear ramp
filter.  For the F622W and F814W observations, IRAS 09104 $+$4109 was
centered on the PC chip.  For the FR680N observations, the galaxy was
positioned on the WF4 chip at a location corresponding to an effective
observed bandpass of $7173-7267$\AA~ ($4974-5039$\AA~ in the rest frame).
At the redshift of IRAS 09104$+$4109, the centers of the F622W and F814W
bandpasses are close to restframe B, and V, respectively, and the FR680N
filter isolates the [OIII] 5007\AA~ emission line.  The F622W data
consisted of four 1000 second integrations and two 900 second
integrations, with the galaxy moved by approximately five PC pixels
between successive exposures.  The F814W data consisted of four 1000
second integrations, with similar telescope movements between exposures.
The FR680N data consisted of two 1300 second integrations. All magnitudes
quoted below are on the Vega system.

After standard WFPC2 pipeline processing, all images in a given filter
were spatially registered and then combined using the cosmic ray rejection
algorithm CRREJ in the NOAO's IRAF reduction package.  The FR680N data
were flattened using a WF4 flat field image taken with the F673N,
narrow--band filter.  A pure [OIII] 5007\AA~ emission--line image was
constructed by subtracting a resampled, scaled (in flux) version of the
F814W PC image, from the FR680N image, after using the nearby early--type
(line--free) galaxies to determine the proper multiplicative flux scale
factor.

\section{Results \& Discussion}

We detect a great deal of structure in the WFPC2 data on sub--kpc scales
in the stellar envelope and circum--nuclear regions of IRAS 09104$+$4109.
The emission--line ``plume" previously seen in [OIII] by Crawford \&
Vanderriest (1996) is resolved into a number of loops and filaments
extending over the entire length of the structure ($\sim5''$ or 25 kpc --
see Figs. 1 and 2).  There is an apparent bend, or break, in the plume
approximately $2''$ from the nucleus, and a possible counter--plume at a
position angle of about $210^{o}$.  The plume itself is responsible for
approximately $10\%$ of the total [OIII] 5007\AA~ emission--line flux from
IRAS 09104$+$4109 (f$_{tot}=1.2\times10^{-13}$ erg cm$^{-2}$ s$^{-1}$).
Note that the plume seen in the F814W image (Fig. 1 \& Fig. 2, {\it top})
is entirely consistent with the expected amount of [OIII] emission within
the filter bandpass.  The primary emission--line plume sits within the
scattering cone recently seen in polarized light by \cite{h99}, but
approximately $45^{o}-50^{o}$ east of the radio jet axis (see Fig. 2, {\it
bottom}).  The northern radio jet sits on the western edge of the
scattering cone.  Since the plume is at the systemic velocity of IRAS
09104$+$4109 and is ionized by a source of hard photons (Crawford \&
Vanderriest 1996), it is likely that this structure represents gas which
is lit up by the mis-directed quasar, and not a radially expanding jet.

Besides the emission--line plume, there are other interesting structures
in the IRAS 09104$+$4109 WFPC2 data.  A number of narrow filaments radiate
from the nucleus along a direction nearly perpendicular to the plume axis
(see Fig. 2, {\it top}).  These ``whiskers" have lengths of $5-7$ kpc,
relatively blue colors (m$_{814}-$m$_{622}=0.7-0.8$ mag, or $\sim0.3-0.5$
mag bluer than the stellar halo), and are not visible in the [OIII]
image.  Since these whiskers radiate from the nucleus along directions
which are apparently shielded from direct quasar light, they may be
associated with gas which is cooling out of the cluster potential and
raining down on the nucleus.  Alternatively, they may be lit up by the
buried quasar through ``cracks" in the obscuring torus.

A second set of structures seen in the IRAS 09104$+$4109 data, are blue
(m$_{814}-$m$_{622}\sim0.5$ mag) filaments at radii of $50-60$ kpc (see
Fig. 2, {\it top}).  These objects range in size from $1''-4''$, or $5-20$
kpc.  The nature of these sources is unknown, but they may represent the
shredded ISM of late--type cluster members.  However, there still appears
to be no obvious evidence for strong galaxy--galaxy interactions in the
recent past between one or more gas--rich galaxies around IRAS
09104$+$4109.

Finally, there are a small number of spheroids within $3''-15''$ of IRAS
09104$+$4109 which are not seen in the ground--based data, and which may
be the bulges of cannibalized cluster galaxies.  These spheroids have
m$_{622}\sim24.5-25.5$ mag, m$_{814}-$m$_{622}\sim1.0-1.5$ mag, and are
all marginally resolved in the PC images, with FWHM $\sim3.5-4.5$ pixels
($\sim1$ kpc).  These objects are clearly not single, proto--globular
clusters, since they are resolved and have high luminosities (the faintest
have M$_{V}\sim -17$ mag -- comparable to a early--type galaxy like M32).
For comparison, the brightest elliptical galaxies in the PC field of IRAS
09104$+$4109 have M$_{V}\sim-21$ mag.  In general, the spheroids
surrounding IRAS 09104$+$4109 have the colors and luminosities expected of
either (1) star--forming clusters with ages of $10^{7}-10^{8}$ yrs and
star formation rates of $\sim10$ M$_{\odot}$ yr$^{-1}$, seen behind $2-3$
magnitudes of visual extinction, or (2) massive ($10^{8}$ M$_{\odot}$),
aging clusters observed at least $10^{7}-10^{8}$ yrs after formation, seen
behind approximately one magnitude of visual extinction (Leitherer \&
Heckman 1995).

Clearly, IRAS09104$+$4109 has a complex structure down to the resolution
limit of the HST.  While we can currently only speculate as to the
relationships between the various stellar, ionized gas, and non--thermal
radio components, high--resolution spectra (i.e. with STIS) should allow
us to better fit this unique galaxy into the current ULIRG framework, and
understand its relationship to other active galaxies as well.

\begin{figure}

\plotfiddle{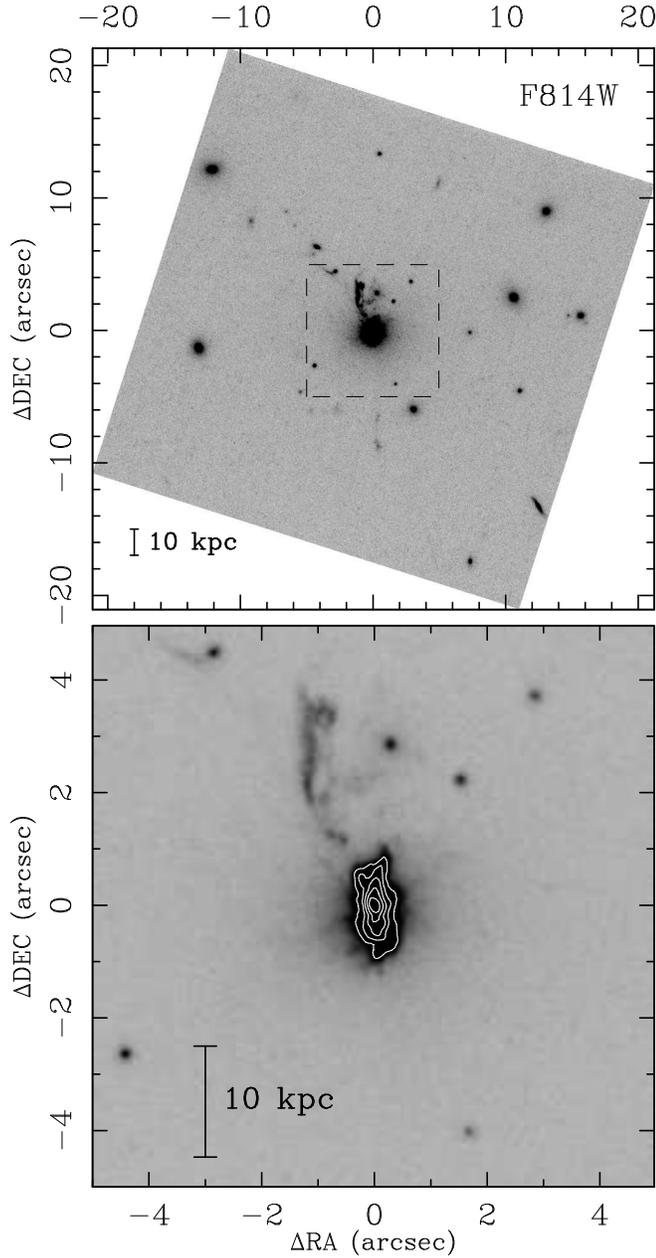}{6.0truein}{0}{84}{84}{-50}{-592}

\vspace{1.55cm}

\caption{Planetary Camera images of IRAS 09104$+$4109 in the F814W
filter.  {\it Top}: the central $\sim40''$ surrounding IRAS 09104$+$4109
(the dashed lines indicate the boundaries of the magnified image shown
below).  {\it Bottom}:  The central $\sim10''$ of the F814W image.
Contours decrease by factors of two, starting at $50\%$ of the peak
value.  In both images, north is up, east is to the left, and a projected
linear scale of 10 kpc is indicated.  The plume is primarily [OIII]
5007\AA~ emission within the F814W bandpass -- see Fig.2}

\end{figure}

\begin{figure}

\plotfiddle{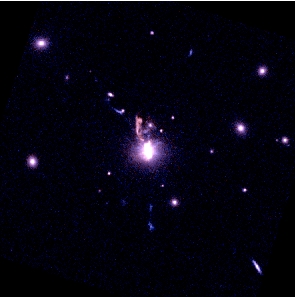}{6.0truein}{0}{82}{82}{-85}{-442}

\plotfiddle{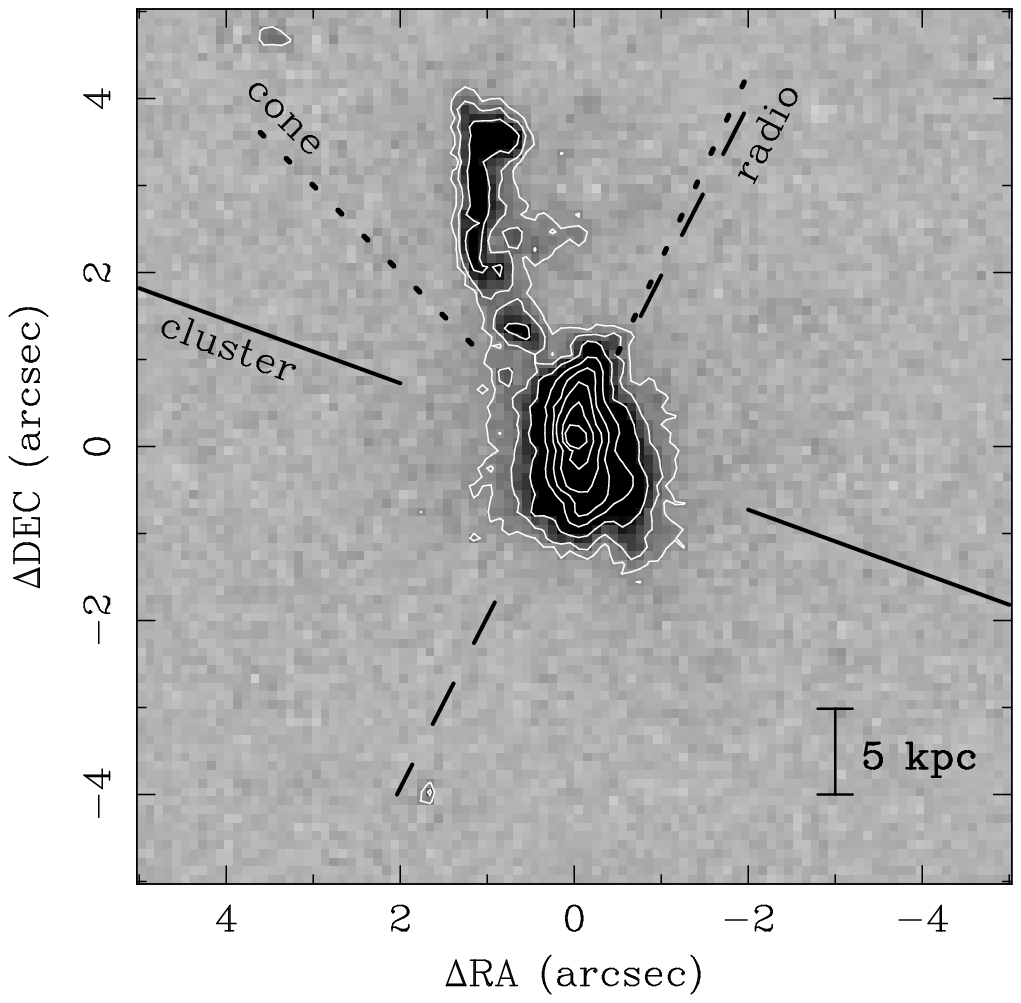}{1.0truein}{0}{79}{79}{-95}{-185}
 
\vspace{-1.38cm}

\caption{{\it Top}: Color image of central $33''$ of the IRAS 09104$+$4109
field constructed from the F622W (blue) and F814W (red) PC images.  North
is up and east is to the left.  Besides the emission--line ``plume"
extending north from the nucleus, note the faint, blue ``filaments" to the
south and north, and the ``whiskers" extending east and west of the
nucleus.  {\it Bottom}: Continuum--subtracted, [OIII] 5007\AA~ image of
IRAS 09104$+$4109, schematically showing the radio (dashed), cluster
(solid) and ionization cone (dotted) axes.  Contours decrease by factors
of two, starting at $50\%$ of the peak value.}

\end{figure}

\end{document}